# Energy Efficient PON Backhaul Network for a VLC Based Fog Architecture

Wafaa B. M. Fadlelmula
School of Electronic and Electrical Engineering
University of Leeds
Leeds, United Kingdom
elwbf@leeds.ac.uk

Sanaa H. Mohamed, *Member IEEE*
School of Electronic and Electrical Engineering
University of Leeds
Leeds, United Kingdom
S.H.H.Mohamed@leeds.ac.uk

Taisir E.H. El-Gorashi
School of Electronic and Electrical Engineering
University of Leeds
Leeds, United Kingdom
T.E.H.Elgorashi@leeds.ac.uk

Jaafar M. H. Elmirghani, *Fellow IEEE*
School of Electronic and Electrical Engineering
University of Leeds
Leeds, United Kingdom
J.M.H.Elmirghani@leeds.ac.uk

*Abstract*— In this paper, an energy efficient passive optical network (PON) architecture is proposed for backhaul connectivity in indoor visible light communication (VLC) systems. The proposed network is used to support a fog computing architecture allowing users with processing demands to access dedicated fog nodes and idle processing resources in other user devices within a building. The fog resources within a building complement fog nodes at higher layers of the access and metro networks and the central cloud data center. A mixed integer linear programming (MILP) model is developed to minimize the total power consumption associated with serving demands over the proposed architecture. The results show that the PON backhaul network improves the energy efficiency of fog computing by 66% for networking and 12% for processing compared to an architecture based on state-of-the-art Spine-and-Leaf connectivity.

## I. Introduction

We are witnessing an unpresented number of devices being connected to the Internet. Based on Cisco Annual Internet Report (2018–2023) [1], 29.3 billion devices will be connected to the Internet by 2023. This increase will be associated with a demand for high date rates and timeliness exceeding the capabilities of the emerging 5G networks. The 6G networks vision promises increased data rates by further exploitation of the electromagnetic spectrum. Optical wireless frequencies offer a potential bandwidth exceeding 540 THz that can complement the Radio Frequency (RF) spectrum in access networks. Several studies have proposed methods to improve the achievable data rates of optical wireless communication (OWC) systems including beam adaptation of the power, angle and delay [2]–[12]. VLC is one of the promising OWC systems that uses (LEDs or LDs) for indoor lighting and communication. For indoor applications, VLC provides high data rates of 25 Gbps and beyond [13]–[24] and enhanced security since the light does not penetrate walls. VLC can also provide low-cost communication as it remains unregulated and unlicensed. Furthermore, VLC is an energy and cost efficient technology as existing lighting infrastructure can be used for communication.

The exponential growth in traffic and processing demands is accompanied by an increase in power consumption. Network energy efficiency has been investigated extensively in the literature including proposing energy efficient architectures for data centres and core nodes, virtualization, integration of renewable energy sources, and content distribution optimization [25]–[54]. In the access network, passive optical networks (PONs) have proven their efficiency in reliably supporting high data rates at low power consumption. PONs have been proposed to provide backhaul connectivity in 5G networks between the radio base stations and the network gateway [55], [56]. PONs have also been proposed for data center interconnection (i.e. inter-rack communication and intra-rack communication) in [57]–[59]. Furthermore PONs have the potential to improve the energy efficiency of fog computing [60], where 75% of the processing will be performed by 2025 [61].

In this paper, we investigate the use of PONs to provide backhaul connectivity for the VLC based fog architecture proposed in [15]. We develop a mixed integer linear programming (MILP) model to minimize the total power consumption of serving demands over the proposed architecture.

The reminder of this paper is organized as follows. Section II presents the PON backhaul network for VLC based fog and introduces the energy efficient resource allocation MILP. Section III, introduces and analyses the MILP model results. Finally the paper is concluded in Section IV.

## II. PON Backhault Network for VLC based Fog Computing architectures

### A. Network architure

In this work, we consider the fog architecture studied in [26]. As shown in Figure 1, each room has a number of VLC access points (APs) serving users. As in [22] all access points use the red wavelength and each access point provides a data rate of 2.5 Gbps for each user. Note that for the shade of white illumination (in VLC) selected during day time, the red colour dominates and hence the red colour offers a higher transmit power, hence its selection in the optimization [22].

The fog computing architecture is composed of idle processing resources in user devices, fog server in each room, and fog nodes at higher layers of the access and metro networks; a building fog node, a campus fog node and a metro fog. Processing resources are also available at the cloud data center.

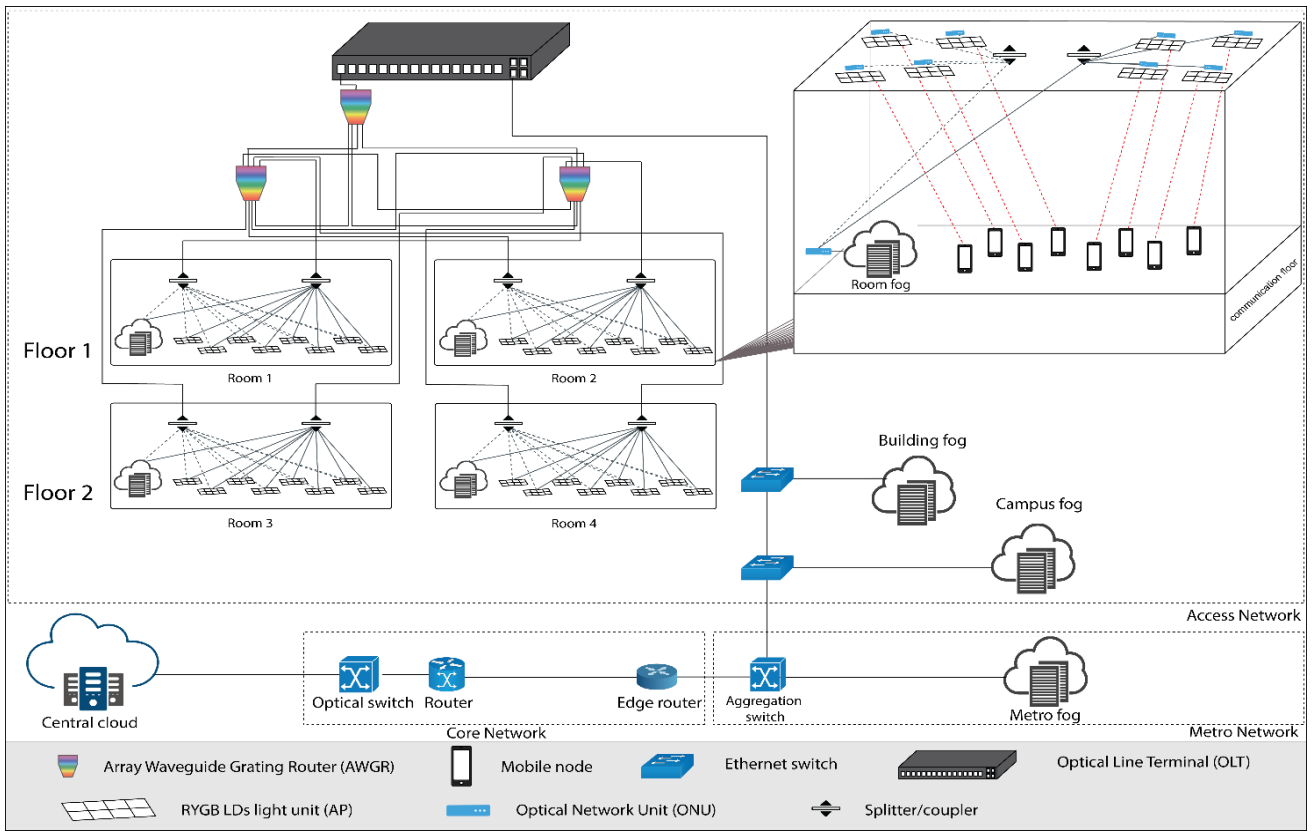

*Figure 1: The fog architecture*

In this work, we consider a PON based architecture adopted from the PON design in [12]. Each access point is connected to an ONU. ONUs within a room are connected to splitters and couplers for upstream and downstream communication, respectively. Two 4×4 arrayed waveguide grating routers (AWGRs) are used to provide connectivity between the rooms. The splitters and couplers of each room are connected to an AWGR input port and an AWGR output port, respectively. The AWGRs facilitate connectivity between APs within a room or in different rooms. The AWGRs are connected to an OLT port to connect the building access network to higher layers. Five wavelengths are used to provide the communication within the access network.

A distinct wavelength is used for communication between the access points in the same room. Additionally, each room has a unique wavelength to transmit the traffic to the OLT. To provide the connectivity between the rooms, three different wavelengths are used. To initiate a connection, the access point sends a control message to the OLT through a specific wavelength. The OLT then grants a dedicated wavelength for the connection.

In Section III, the PON backhaul network is compared to a spine-and-leaf [62] based backhaul as shown is Figure 2. The access points are connected to leaf switches which are connected to spine switches and a router. The router connects the access network to the higher layers through an Ethernet switch.

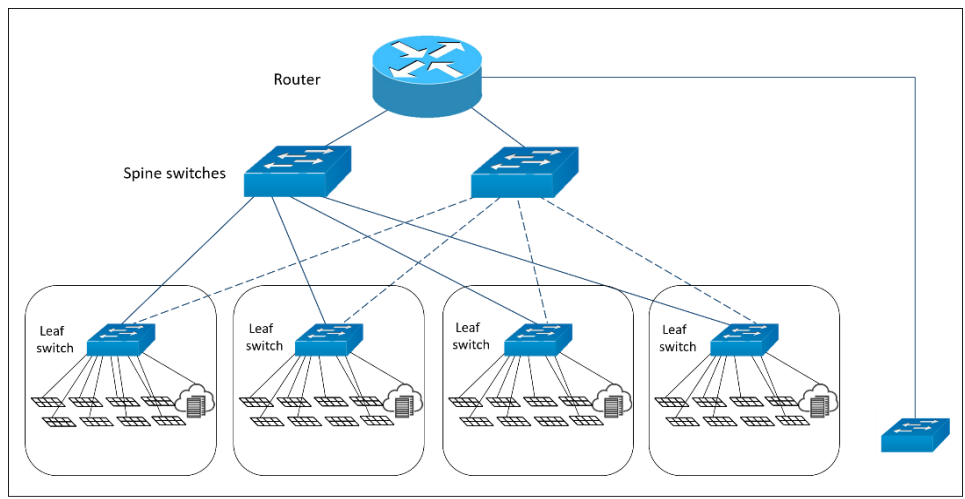

*Figure 2: spine-and-leaf architecture*

## B. Energy Efficient Resource Allocation MILP model:

We developed a MILP model to optimally allocate the processing resources to serve demands for minimum processing and networking power consumption. The objective function of the model is given by:

$$\text{Minimize:} \quad TPC = P_p + P_n \quad (1)$$

where $TPC$ is the total power consumption, $P_p$ is the processing power consumption of the processing nodes (i.e., user devices, room fog, building fog, campus fog, metro fog and the cloud), and $P_n$ is the power consumption of devices traversed by traffic on the network. All processing nodes and networking devices are considered to follow a linear power profile with minimum power consumption to activate the device regardless of the device workload.

The model is subject to the following constraints:

$$\sum_{d \in P} \psi^{sdk} = D_{sk}, \forall s \in M, k \in K_s \quad (2)$$

Constraint (2) is used to ensure that the processing resources allocated to a processing node meet the demand, where the set of mobiles generating the demand, the set of tasks and the set of processing nodes are defined as $M, K_s, P$ respectively. $\psi^{sdk}$ is the total processing assignment in the processing node $d$, generated by user device $s$ for the task $k$; $D_{sk}$ is the processing demand generated by mobile $s$ requesting task $k$.

$$\sum_{s \in M} \sum_{k \in K_s} \psi^{sdk} \leq \Omega_d, \forall d \in P. \quad (3)$$

Constraint (3) ensures that the processing allocated to a node does not exceed its capacity, where $\Omega_d$ is the processing capacity of the processing node $d$.

$$\sum_{s \in M, d \in P, k \in K_s} \lambda_{ijw}^{sdk} \leq L_{ijw}, \quad (4)$$
$$\forall i \in N, j \in N_i, w \in W, i \neq j.$$

Constraint (4) ensures that the traffic in the link between the source and destination using wavelength $j$ does not surpass the capacity of that link, where $N$ is the set of all nodes, while $W$ is the set of wavelengths. $\lambda_{ijw}^{sdk}$ is the traffic flow between source and destination pair $(s, d)$ for task $k$ traversing the physical link $(i, j)$, that uses the wavelength $w$; and $L_{ijw}$ is the total capacity of the link between the physical nodes $(i, j)$ using a wavelength $w$.

In addition to these constraints, the model is subject to the traffic flow conservation constraint, a set of constraints to ensure wavelength continuity in connections between source and destination pairs, and a constraint to ensure that each task is assigned to one processing node only.

## III. RESULTS AND DISCUSSION

In this section, we examine a scenario of a four room building. Each room has eight VLC access points, each connecting users. Half of the users in each room generate demands while the rest offer their resources to the fog computing architecture. Each user generates a single demand. The demands processing load takes values in the range 100 MIPS-1500 MIPS. The traffic demand is related to the processing demand by the Data Rate Ratios (DDR) (the ratio of the traffic demand in MIPS to the processing demand in Mbps). To represent applications with intensive processing and communication demands, a DDR of 0.6 is considered (i.e. the traffic demands take values in the range 60-900 Mbps). Table 1 and Table 2 summarise the parameters of the processing nodes and networking devices, respectively.

TABLE 1: PROCESSING DEVICES PARAMETERS

| Parameter | Value |
|---|---|
| Processing capacity of the Mobile device | 1500 MIPS [63] |
| Processing capacity of the Room fog server | 5000 MIPS [64] |
| Processing capacity of the building fog server | 14000 MIPS [65] |
| Processing capacity of the campus fog server | 35160 MIPS [66] |
| Processing capacity of the metro fog server | 73440 MIPS [67] |
| Processing capacity of the cloud server | 320440 MIPS [68] |
| Maximum power consumption of each mobile device | 6.6 Watts [63] |
| Maximum power consumption of the room fog server | 15 Watts [64] |
| Maximum power consumption of the building fog server, Campus fog server, Metro fog server | 95 Watts [65]–[67] |
| Maximum power consumption of the cloud server | 120 Watts [68] |
| Dara rate ratio (DDR). | 0.6 |

TABLE 2: NETWPRKING DEVICES PARAMTERS

| Network device | Maximum power consumption (Watts) | Idle power consumption (Watts) | Capacity (Mbps) |
|---|---|---|---|
| Access Point | 7.2 [22] | 6.48 | 2500 [22] |
| ONU | 15 [69] | 13.5 | 10000 [69] |
| OLT Line card | 300 [70] | 270 | 160000 [70] |
| Ethernet switch | 3800 [71] | 3420 | 160000 [71] |
| Aggregation switch | 3800 [71] | 3420 | 160000 [71] |
| Edge router | 4200 [72] | 3780 | 200000 [72] |
| Optical switch | 63.2 [73] | 56.88 | 100000 [73] |
| Core router | 13200 [74] | 11880 | 1200000 [74] |

The performance of the PON backhaul network is compared to a spine-and-leaf backhaul based network. Table 3 summarizes the values of the parameters of the spine-and-leaf network.

TABLE 3: PARAMETER FOR SPINE-AND-LEAF ARCHITECTURE

| Network device | Maximum power consumption (watts) | Idle power consumption (watts) | Capacity (Mbps) |
|---|---|---|---|
| Spine-and-leaf switches | 193 [75] | 173.7 | 240000 [75] |
| router | 4200 [72] | 3780 | 200000 [72] |

Figure 4.a and Figure 4.b show the allocation of fog and cloud resources to processing demands of different workloads in the PON-based architecture and the spine-and-leaf

architecture, respectively. Figure 5.a and Figure 5.b show the total power consumption broken down into processing and networking power consumption resulting from the resource allocation in Figure 4.a and Figure 4.b, respectively.

As observed in Figure 4.a, the MILP power minimization optimization results show that for demands of 100 MIPS, all the demands are served in a single room fog server (room fog 3). As the demand increases and the room fog server reaches its capacity, another room fog is activated before accessing any further remote fog units in the access or metro for example. The same network power consumption results from accessing the room fogs and accessing the user devices passive network. However, serving the demands from the room fogs is more efficient in terms of processing. At 1300 MIPS, all the room fogs capacity is exhausted, and hence the idle user devices are activated to serve the demands in addition to the room fogs. The user devices are less efficient in terms of processing. Therefore, they are activated after the room fogs. The demands considered did not require the activation of higher fog nodes and cloud resources.

The spine-and-leaf backhaul network shows a different trend in allocating resources in Figure 4.b. The demands of each room are served from the fog server located in the same room. Serving all the demands in one room fog will lead to a significant increase in the network power consumption due to the need to go through a second layer of spine switches. As a consequence of avoiding the spine switches, the power consumption of the processing will increase as four servers are activated compared to one in the case of the PON-based architecture. Similar to the PON backhaul network, the mobile devices are activated to serve demands of 1300 MIPS and higher.

Comparing the power consumption levels in Figure 5.a and Figure 5.b shows that 66% of the networking power consumption and 12% of the processing can be saved by deploying the PON backhaul network compared to the spine-and-leaf architecture.

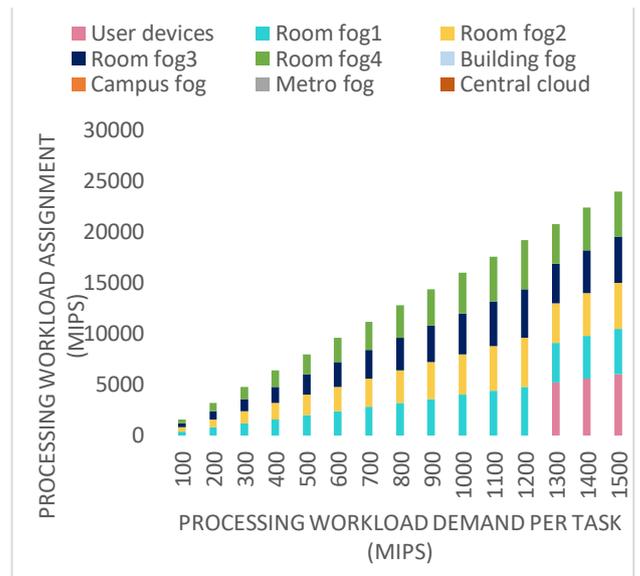

*Figure 4.b: Processing workload allocation in a spine-and-Leaf architecture*

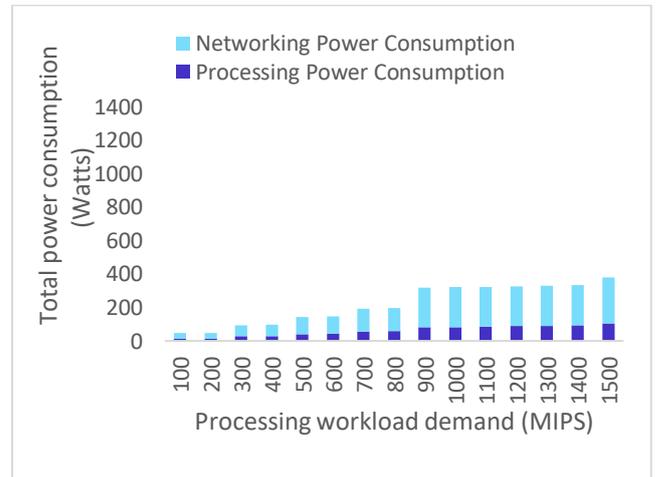

*Figure 5.a: Power consumption of networking and processing in a PON based architecture.*

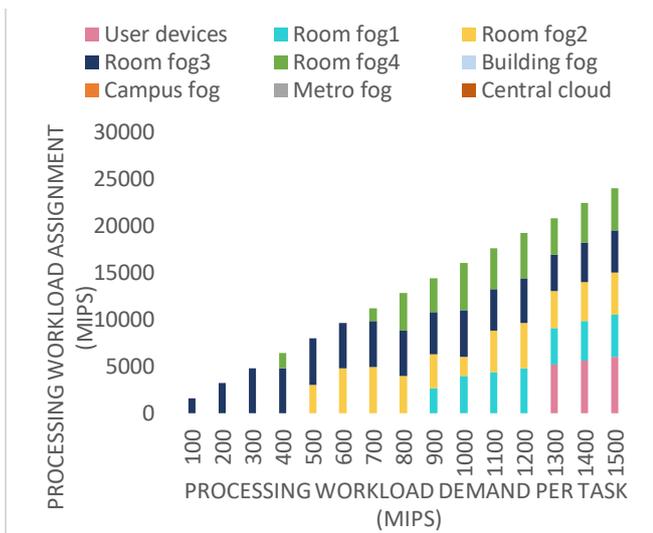

*Figure 3.a: Processing workload allocation in a PON based architecture.*

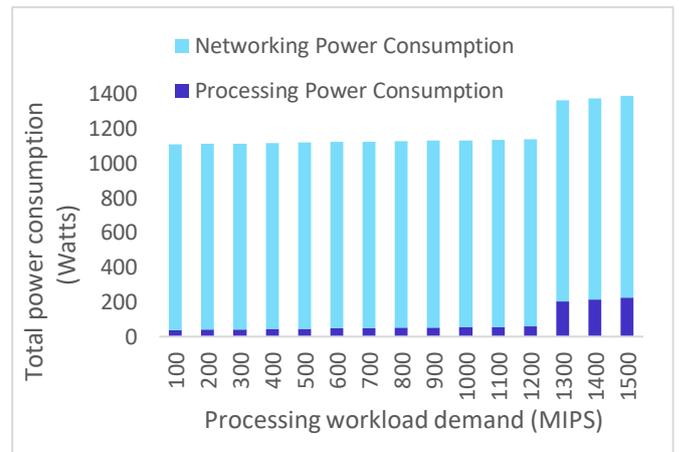

*Figure 5.b: Power consumption of processing and networking in spine-and-Leaf architecture.*

## IV. CONCLUSIONS AND FUTURE WORK

This paper proposed an energy efficient PON backhaul network for a VLC based fog architecture where fog resources within a building complement fog nodes at higher layers of the access and metro networks and the central cloud data center. We developed a mixed integer linear programming (MILP) model to minimize the total power consumption of serving demands. The resource allocation results show the ability of the PON backhaul network to consolidate demands in fewer nodes resulting in improvements in the energy efficiency of the fog architecture by 66% for networking and 12% for processing compared to an architecture based on a Spine-and-Leaf based backhaul.

ACKNOWLEDGMENTS

The authors would like to acknowledge funding from the Engineering and Physical Sciences Research Council (EPSRC) INTERNET (EP/H040536/1), STAR (EP/K016873/1) and TOWS (EP/S016570/1) projects. All data are provided in full in the results section of this paper.